\documentclass[10pt]{iopart}
\usepackage{graphicx,epsfig}
\begin{document}
\title{Acceleration disturbances and requirements for ASTROD I}
\author{Sachie Shiomi$^{1}$ and Wei-Tou Ni$^{1, 2, 3}$}%
\address{$^{1}$Department of Physics, National Tsing Hua University,
Hsinchu, Taiwan 30013 ROC\\ $^{2}$Purple Mountain Observatory,
Chinese Academy of Sciences, Nanjing 210008 China\\ $^{3}$National
Astronomical Observatories, Chinese Academy of Sciences, Beijing
100012 China}
\begin{abstract}

ASTRODynamical Space Test of Relativity using Optical Devices I
(ASTROD I) mainly aims at testing relativistic gravity and measuring
the solar-system parameters with high precision, by carrying out
laser ranging between a spacecraft in a solar orbit and ground
stations. In order to achieve these goals, the magnitude of the
total acceleration disturbance of the proof mass has to be less than
$10^{-13}$\,m\,s$^{-2}$\,Hz$^{-1/2}$ at 0.1\,m\,Hz. In this paper,
we give a preliminary overview of the sources and magnitude of
acceleration disturbances that could arise in the ASTROD I proof
mass. Based on the estimates of the acceleration disturbances and by
assuming a simple control-loop model, we infer requirements for
ASTROD I. Our estimates show that most of the requirements for
ASTROD I can be relaxed in comparison with Laser Interferometer
Space Antenna (LISA).

\end{abstract}
\pacs{04.80.Nn, 04.80.-y, 95.55.Ym}

\section{Introduction}
ASTRODynamical Space Test of Relativity using Optical Devices
(ASTROD) \cite{ASTROD1, ASTROD2} aims at testing relativistic
gravity, measuring the solar-system parameters with high precision
and detecting gravitational waves from massive black holes and
galactic binary stars. The concept of ASTROD is to put two
spacecraft in separate solar orbits and carry out laser
interferometic ranging with Earth reference stations (e.g. a
spacecraft at the Earth-Sun L1/L2 points). A simple version of
ASTROD, ASTROD I\footnote{ASTROD I was previously referred to as
Mini-ASTROD (for instance in \cite{Ni2002}).}, has been studied as
the first step to ASTROD \cite{Ni2002,Ni2004}. ASTROD I employs one
spacecraft in a solar orbit and carries out interferometric ranging
and pulse ranging with ground stations. The main scientific goals of
ASTROD I are to test relativistic gravity and the fundamental laws
of spacetime with three-order-of-magnitude improvement in
sensitivity and to improve the solar, planetary and asteroid
parameter determination by 1 to 3 orders of magnitude. The
technological goal of ASTROD I is to prepare for the ASTROD mission.

\begin{figure}[htbp]
    \centering
\psfig{file=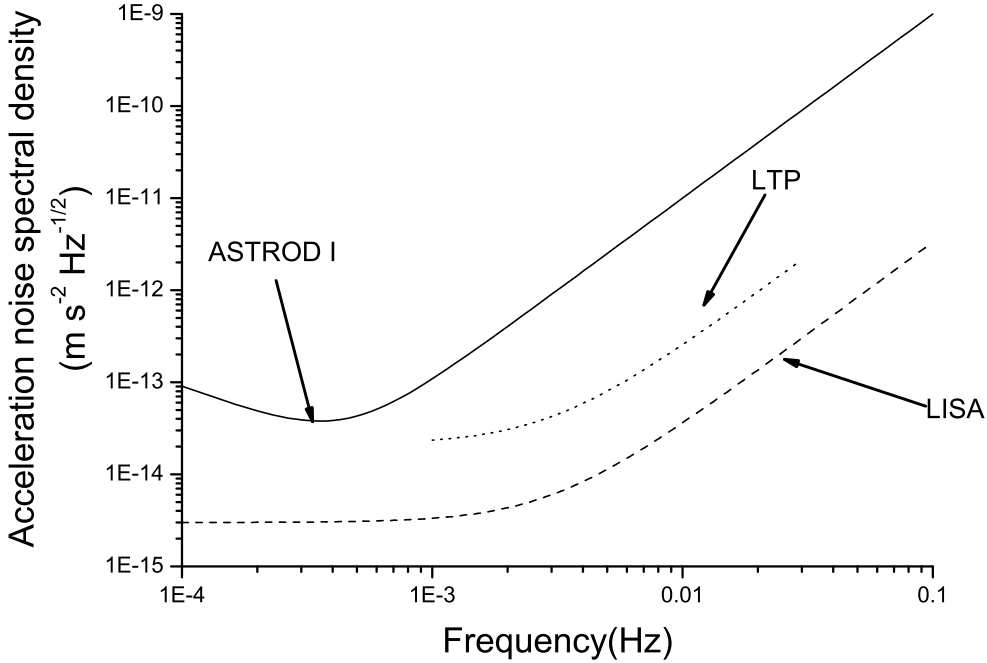,width=\linewidth} \caption{Acceleration
noise spectral density requirements for ASTROD I. The LISA
Technology Package (LTP) requirements \cite{Anza2005} and LISA
requirements \cite{LISA} are illustrated in the figure for
comparison.} \label{Fig:Curve}
\end{figure}

The acceleration disturbance goal of the ASTROD I proof mass is
$10^{-13}$\,m\,s$^{-2}$\,Hz$^{-1/2}$ at frequency $\nu$ of 0.1\,mHz.
The power spectral density of the allowed level of the acceleration
noise is shown in figure \ref{Fig:Curve}. Assuming a 10\,ps one-way
timing accuracy (3\,mm ranging accuracy) and the acceleration noise
of $10^{-13}$\,m\,s$^{-2}$\,Hz$^{-1/2}$ at frequency of about
0.1\,mHz, a simulation for 400 days (350--750 days after launch)
showed that ASTROD~I could determine the relativistic parameters
$\gamma$ and $\beta$, and the solar quadrupole parameter $J_2$ to
levels of 10$^{-7}$, 10$^{-7}$ and 10$^{-8}$, respectively
\cite{Tang2004}. In the simulation, (i) the timing noise is modeled
as Gaussian random noise; (ii) unknown acceleration noise is modeled
to have Gaussian random magnitude with zero mean and with standard
deviation 10$^{-15}$\,m\,s$^{-2}$ and to have its direction changed
randomly every 4\,h (equivalent to
10$^{-13}$\,m\,s$^{-2}$\,Hz$^{-1/2}$ for $\nu$ $\sim$ 0.1\,mHz
assumed as the requirement of the drag-free system) and (iii) five
range points are taken each day (at 0.2\,d interval). Longer term
systematic effects will be studied in a future paper. This
simulation agrees with the scientific goals of ASTROD~I. The timing
uncertainty of event timer reaches 3\,ps in satellite laser ranging
at present. Space qualified versions of similar accuracy are under
development. For a ranging uncertainty of 3\,mm in a distance of 3
$\times$ 10$^{11}$\,m (2\,AU), the laser/clock frequency needs to be
known to one part in 10$^{14}$. This can be set as a requirement of
the space laser/clock or a requirement for laser frequency
monitoring through ground clock and modeling. As to ground station
jitter, monitoring to an accuracy of 3\,mm is required and can be
achieved. The atmospheric effects on laser propagation will be
monitored and subtracted to mm-level by using 2-color (2-wavelength)
ranging (one color for pulse ranging and one for interferometric
ranging). These measurement uncertainties are not cumulative in the
range determination while the acceleration disturbances accumulate
in time in the geodesic deviations. In order to achieve the
acceleration disturbance goal, a drag-free control system using
capacitive sensors will be employed.

In order to decide on detailed designs of the accelerometer for
ASTROD I, we have to know the sources and magnitude of acceleration
disturbances that could arise in the accelerometer. In this paper,
we carry out analyses, mainly based on existing literature on
acceleration disturbances for other gravitational missions (LISA
\cite{LISA,FTR} and the LISA Pathfinder \cite{Anza2005,LTP}), to
give a preliminary overview of the acceleration disturbance to the
ASTROD I proof mass. Based on the analyses, we infer some of the
requirements for the designs of the ASTROD I payload and spacecraft.
Also, we compare parameter values we have assumed in the analyses
with those for LISA to confirm the feasibility of the requirements
for ASTROD I.

First, we will give an overview of the ASTROD I configuration
(section \ref{configuration}) and the control-loop model we assumed
(section \ref{control-loop}). Then, we will estimate the magnitude
of acceleration disturbances and requirements for ASTROD I in
sections \ref{Ans} to \ref{st:requirements}, and compare the
requirements of ASTROD I with LISA in section \ref{comparison with
LISA}.

\section{ASTROD I spacecraft configuration}\label{configuration}

The ASTROD I spacecraft has a cylindrical shape with diameter 2.5\,m
and height 2\,m. Its cylindrical side is covered by solar panels.
The cylindrical axis is perpendicular to the orbit plane and the
telescope is set to point toward a ground laser station. The
effective area of receiving sunlight is about 5\,m$^2$ and it can
generate power that is larger than 500\,W. The total mass of the
spacecraft is about 350\,kg and that of payload is 100--120\,kg (see
\cite{Ni2004,Ni-Shiomi2004} for more detailed descriptions of the
configuration). The orbit distance from the Sun varies from about
0.5\,AU to 1\,AU (figure 2 of \cite{Ni-Shiomi2004}).

The proof mass ($m_p$ = 1.75\,kg) is a rectangular parallelepiped
(50 $\times$ 50 $\times$ 35\,mm$^3$)\footnote{This is the current
design of the proof mass. A cylindrical shape is also considered as
an alternative design of the proof mass.} made from Au-Pt alloy
(density $\rho$ = 2 $\times$ 10$^4$\,kg\,m$^{-3}$). The six sides of
the proof mass are surrounded by electrodes mounted on the housing
for capacitive sensing and control. The gap between each side of the
proof mass and the opposing electrode is 2\,mm. Assumed values for
the capacitance and voltages are listed in table
\ref{Table:parameter values}.

\section{Control-loop model}\label{control-loop}

\begin{figure}
    \centering
\psfig{file = 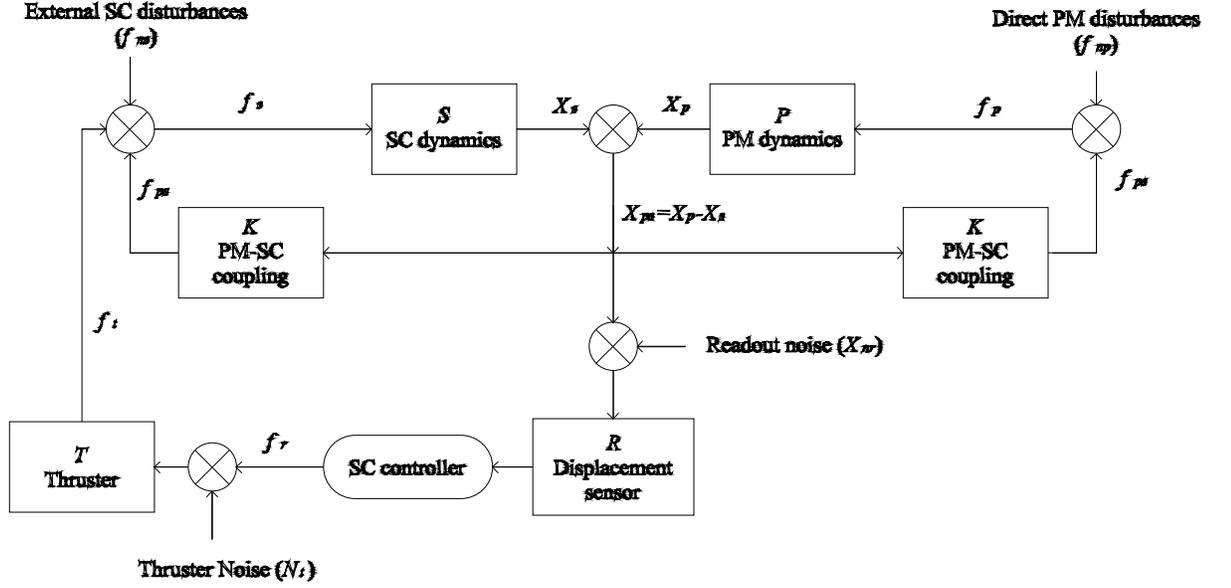, width=18 cm} \caption{Control
loop model. SC and PM denote spacecraft and proof mass,
respectively. }\label{Fig:control-loop}
\end{figure}

Various acceleration disturbances would act on the proof mass in
different ways. In order to infer how these different kinds of
acceleration disturbances would contribute to the total acceleration
disturbance of the proof mass, we tentatively assume a simple
single-mass and single-axis control-loop model
\cite{Vitale2000,Vitale2002}. The diagram is shown in
figure~\ref{Fig:control-loop}. The relative difference between the
displacement disturbance amplitudes of the proof mass and of the
spacecraft, $X_{ps} = X_p - X_s$, with position readout noise
$X_{nr}$, is measured by the position displacement sensor. The
output of this sensor is converted to acceleration disturbance
$f_{r}$, by a transfer function $R$ ($\equiv \omega_R^2$). This
acceleration is supplied to the thruster and the output acceleration
disturbance with thruster noise $N_t$ is applied to the spacecraft.
The spacecraft also experiences acceleration disturbances by
coupling to the proof mass with a coupling constant $K$ and external
environmental disturbances ($f_{ns}$) that work directly on the
spacecraft. The total acceleration of the spacecraft $f_s$ is
converted to the position noise $X_{s}$ with a transfer function
$S$.

We consider only the sensitive axis of the proof mass without
actuation. The proof mass would experience disturbances by
spacecraft-proof mass coupling and environmental disturbances
$f_{np}$ (see section \ref{Anp}). The total acceleration disturbance
of the proof mass $f_p$ is converted to the displacement disturbance
$X_p$ with a transfer function $P$.

From this control-loop model, we obtain the following linear loop
equations \cite{Schumaker2002,Schumaker2003}:
\begin{eqnarray}
f_r &=& R(X_{ps}+X_{nr})\\ f_t &=& T(f_r + N_t)\\ f_s &=& f_{ns} +
f_t - (m_p/M_{sc})f_{ps}\\ f_p &=& f_{np} +f_{ps}\\ f_{ps} &=&
KX_{ps} \\ X_s &=& Sf_s\\ X_p &=& Pf_p
\end{eqnarray}

By solving these equations for $X_p$, and assuming $S = P =
\omega^{-2}$ and $u$ $\equiv$ $STR$ (= $T \omega_R^2 \omega^{-2}$;
$T$ is nominally 1 and its effects are absorbed in $R$ and $N_t$),
we obtain (e.g.
\cite{Vitale2000,Vitale2002,Schumaker2002,Schumaker2003}):
\begin{equation}
f_p \approx X_{nr}(-K) + f_{np} + (f_{ns} +
TN_t)K\omega^{-2}u^{-1}\label{eq:Control-model}
\end{equation}
where $\omega = 2 \pi \nu$. This acceleration disturbance has to be
less than the acceleration noise goal of $10^{-13}
$\,m\,s$^{-2}$\,Hz$^{-1/2}$ at $\nu$ = 0.1\,mHz.

We will estimate the values of the direct acceleration
disturbances of the spacecraft ($f_{ns}$ and the thruster noise
$TN_t$) in section \ref{Ans} and of the proof mass ($f_{np}$) in
sections \ref{Anp} and \ref{back-action}, and the stiffness $K$ in
section \ref{stiff}. We will discuss the requirements for $X_{nr}$
and $u$ in section \ref{st:requirements}.

\section{Direct acceleration disturbances of the spacecraft}\label{Ans}
The spacecraft would be affected by environmental disturbances that
stem from, for example, solar radiation pressure, solar wind and
micrometeorite impacts. Among these sources of disturbances, solar
radiation pressure is considered to be the major contributor to the
acceleration disturbances (section 7 of \cite{FTR}). The
contribution from solar wind might be comparable to radiation
pressure, but the spectral behavior of the solar wind is not well
known.

By assuming a perfectly reflecting surface of the spacecraft,
acceleration noise caused by fluctuation in solar irradiance $\delta
W_0$ is $f_{ns,srp} = \frac{4}{3}(A_{sc} \delta W_0/M_{sc}c)$, where
$A_{sc}$ is the area of the spacecraft facing the Sun, $M_{sc}$ is
the mass of the spacecraft and $c$ is the speed of light in vacuum.
From the data of the VIRGO experiment on SOHO \cite{Pap1999},
fractional fluctuation in solar irradiance is $\delta W_0/W_0
\approx 2.8 \times 10^{-3}$\,Hz$^{-1/2}$ at 0.1\,mHz (figure 6 of
\cite{Pap1999}). Assuming that $\delta W_0/W_0$ is the same at
0.5\,AU, and taking a total irradiance of 5500\,W\,m$^{-2}$, the
area of 5\,m$^2$ and a mass of 350 kg, we obtain $f_{ns,srp}$ = 9.8
$\times$ 10$^{-10}$\,m\,s$^{-2}$\,Hz$^{-1/2}$.

The impact rate of 1-ng meteorites on the ASTROD I spacecraft (its
surface area is 25.5\,m$^2$) is about 7 events per day or about
0.08\,mHz\footnote{This rate was tentatively estimated from the
meteorite flux/mass distribution used by the LISA study team (figure
7.2-29 of \cite{FTR}).}. The average velocity of meteorites is
18\,km\,s$^{-1}$ in an Earth frame near the Earth (figure 7.2-30 of
\cite{FTR}). An impact of 1-ng meteorite with the average velocity
of 18\,km\,s$^{-1}$ on the surface of the spacecraft (350\,kg)
without reflection produces a linear velocity increment of about 5
$\times$ 10$^{-11}$\,m\,s$^{-1}$. Smaller meteorites have larger
flux (figure 7.2-29 of \cite{FTR}), but their impacts on velocity
changes are smaller. Therefore, the contribution to the acceleration
disturbances at 0.1\,mHz seems insignificant, in comparison with the
effect of solar radiation pressure. However, because such discrete
changes could directly affect the ASTROD~I experiment, the impact
effects have to be studied carefully in detail.

In addition to the acceleration noise from the environmental
disturbances, the spacecraft would suffer from the thruster noise. A
force fluctuation of 10\,$\mu$N\,Hz$^{-1/2}$ in thruster corresponds
to the acceleration disturbance of $TN_t$ = 2.8 $\times$
10$^{-8}$\,m\,s$^{-2}$\,Hz$^{-1/2}$.

Therefore, the rss (root-sum-square) of these acceleration
disturbances is:
\begin{equation}
f_{ns}+TN_t \approx 2.8 \times 10^{-8}
\hspace{5pt}\rm{m}\hspace{5pt}
\rm{s}^{-2}\hspace{5pt}\rm{Hz}^{-1/2}
\end{equation}
at 0.1\,mHz, dominated by the thruster noise. As per \cite{LISA},
LISA requires a force fluctuation to be less than a few
$\mu$N\,Hz$^{-1/2}$. However, recent studies for the LISA Pathfinder
indicate that force noise up to 0.1\,mN\,Hz$^{-1/2}$ at 0.1\,mHz can
be tolerated by increasing the gain \cite{Fichter2005}; the LISA
study team is designing high gain loops and the requirement has been
relaxed\footnote{This is a comment from one of the referees.}.

\section{Direct acceleration disturbances of the proof mass}\label{Anp}

Direct proof-mass acceleration disturbances can be classified into
two categories depending on their origins
\cite{Schumaker2002,Schumaker2003}: environmental disturbances
($f_{nep}$) and proof-mass sensor back-action acceleration
disturbances ($f_{nbp}$). The former includes disturbances related
to magnetic effects ($f_{m1}$, $f_{m2}$, $f_{m3}$, $f_{L1}$ and
$f_{L2}$), impact effects (by cosmic ray ($f_{c}$) and residual gas
($f_{rg}$)), temperature dependent effects (radiometric and
outgassing effects ($f_{re}$ and $f_{og}$) and thermal radiation
pressure ($f_{tr}$)) and gravity gradients caused by thermal
distortion of the spacecraft ($f_{gg}$). The latter is originated
from voltage fluctuations ($f_{b1}$ and $f_{b2}$), charge
fluctuations ($f_{b3}$ and $f_{b4}$), readout electronics
($f_{ba}$), patch field voltages ($f_{pe}$) and thermal voltage
noise by dielectric losses ($f_{dl}$).

Parameter values and physical constants used for the estimations are
listed in tables \ref{Table:parameter values} and
\ref{Table:Physical constants}, respectively. Table \ref{Table:PM
acceleration disturbances} provides a summary of the expressions
used to estimate the direct proof-mass acceleration disturbances and
the estimated values. The disturbances noted as $f_{m1}$, $f_{m2}$,
$f_{L1}$, $f_{L2}$, $f_{c}$, $f_{rg}$, $f_{re}$, $f_{tr}$, $f_{gg}$
and $f_{b1}$-$f_{b4}$ in table~\ref{Table:PM acceleration
disturbances} correspond to $A_1$-$A_{6}$ and $A_8$-$A_{14}$ in
\cite{Schumaker2003}, respectively\footnote{The acceleration
disturbance due to laser photon radiation pressure, noted as $A_7$
in \cite{Schumaker2003}, would not arise in ASTROD~I, as the laser
beam is not to be injected directly on the surface of the ASTROD~I
proof mass.}. We will briefly explain each disturbance below.

\begin{table}
    \caption{\label{Table:parameter values}Parameter values used in the
acceleration noise estimates}
    \begin{indented}
    \item[]\begin{tabular}{@{}ll}
  \br
   $\nu$ = 0.1 mHz & ASTROD I \\
   \mr
  \multicolumn{2}{l}{\textbf{Proof Mass (PM)}} \\
  Mass: $m_p$ [kg] & 1.75 \\
  Density: $\rho$ [kg m$^{-3}$] & $2 \times 10^4$ \\
  Cross section: $A_p$ [m$^2$] & 0.050 $\times$ 0.035 \\
  Temperature:  $T_p$ [K] & 293 \\
  Fluctuation of temperature difference & \\
  \hspace{1pt} across PM and housing: $\delta T_d$ [K Hz$^{-1/2}$] & 1.0
$\times$ $10^{-5}$ \\
  Maximum charge build-up: $q$ [C] & $10^{-12}$ \\
  Fluctuation in charge: $\delta q$ [C Hz$^{-1/2}$] & 6.1 $\times$
10$^{-15}$ \\
  Magnetic susceptibility: $\chi_m$ & $5 \times 10^{-5}$ \\
  Magnetic remanent moment: $|\vec{M_r}|$ [A m$^2$] & 1 $\times$ 10$^{-7}$
\\
  Residual gas pressure: $P$ [Pa] & $10^{-5}$ \\
  Electrostatic shielding factors: $\xi_e$ & 10 \\
  Magnetic shielding factors:  $\xi_m$ & 1 \\
  \mr
  \multicolumn{2}{l}{\textbf{Spacecraft (SC)}} \\
  Mass: $M_{sc}$ [kg] & 350 \\
  Velocity: $v$ [m s$^{-1}$] & $4 \times 10^4 $ \\
  Thruster noise [$\mu$N Hz$^{-1/2}$] & 10 \\
  Area facing the Sun: $A_{sc}$ [m$^2$] & 5 \\
  Fluctuation of temperature in SC: $\delta T_{sc}$ [K Hz$^{-1/2}$] & 0.2 \\
  \mr
  \multicolumn{2}{l}{\textbf{Capacitive sensing}}\\
  Capacitance: $C_x$ [pF] & 6 \\
  Capacitance to ground:  $C_g$ [pF] & 6 \\
  Total capacitance: $C$ [pF] & ($6C_g$ $\approx$)36 \\
  Gap: $d$ [mm] & 2 \\
  Asymmetry in gap across opposite sides of PM: $\Delta d$ [$\mu$m] & 10 \\
  Average voltage across opposite faces: $V_{x_0}$ ($\equiv$
  $(V_{x1}+V_{x2})/2$) [V] & 0.1 \\
  Proof mass bias voltage: $V_{M0}$ [V] & 0.6 \\
  Voltage difference between $V_{x_0}$ and voltage to ground: $V_{0g}$ [V] &
0.05 \\
  Voltage difference between opposing faces: $V_d$ [V] & 0.01 \\
  Fluctuation of voltage difference across opposite faces: & \\
  \hspace{1pt} $\delta V_d$ [V Hz$^{-1/2}$] & $10^{-4}$ \\
  Residual dc bias voltage on electrodes: $V_0$ [V] & 0.1 \\
  Loss angle: $\delta$ & 10$^{-5}$ \\
  \mr
  \multicolumn{2}{l}{\textbf{Magnetic fields}} \\
  Local magnetic field: $B_{sc}$ [T] & 8 $\times$ 10$^{-7}$ \\
  Local magnetic field gradient: $|\vec{\nabla} B_{sc}|$ [T m$^{-1}$] & 3
$\times$ 10$^{-6}$ \\
  Fluctuation in local magnetic field: $|\delta B_{sc}|$ [T Hz$^{-1/2}$] & 1
$\times$ $10^{-7}$ \\
  Interplanetary magnetic field: $B_{ip}$ [T] & $1.2 \times 10^{-7}$ \\
  Interplanetary magnetic field gradient: $|\delta B_{ip}|$ [T Hz$^{-1/2}$]
& $4 \times 10^{-7}$ \\
  Gradient of time-varying magnetic field: $|\vec{\nabla}(\delta B)|$ [T
m$^{-1}$ Hz$^{-1/2}$] &4 $\times$ 10$^{-8}$ \\
  \mr
  \multicolumn{2}{l}{\textbf{Cosmic rays}} \\
    Impact rate: $\lambda$ [s$^{-1}$] & 30 \\
    Mass of positron: $m$ [kg] & $1.7 \times 10^{-27}$ \\
    Incident energy: $E_d$ [J] & (200 $\mathrm{MeV}$ =) 3.2 $\times$
10$^{-11}$ \\
    \mr
  \multicolumn{2}{l}{\textbf{Gravity gradients}}\\
    Source mass: $M_{dis}$ [kg] & 1 in $f_{gg}$ and 0.03 in $a_{gg}$ \\
    Distance from the source mass: $x$ [m] & 0.5 in $f_{gg}$ and 0.05 in
$a_{gg}$ \\
    Thermal expansion coefficient: $CTE$(Aluminium) [K$^{-1}$] & 2.5
$\times$ 10$^{-5}$ \\
    \mr
    \multicolumn{2}{l}{\textbf{Patch field}}\\
    Multiplicative factor: $\gamma$ & 5 \\
    Patch field voltage: $V_{pe}$ [V] & 0.1 \\
    \br
  \end{tabular}
  \end{indented}
\end{table}

\begin{table}
  \caption{\label{Table:Physical constants}Physical constants used in the
estimation}
    \begin{indented}
    \item[]\begin{tabular}{@{}lll}
    \br
     & Symbols & Values used in the estimations \\
    \mr
    Elementary charge & $e$ & 1.6 $\times 10^{-19}$ C \\
    Permeability of vacuum & $\mu_0$ & 1.26 $\times 10^{-6}$ N A$^{-2}$  \\
    Permittivity of vacuum & $\varepsilon_0$ & 8.9 $\times$ 10$^{-12}$ F
m$^{-1}$\\
    Speed of light in vacuum & $c$ & 3.0 $\times 10^{8}$ m s$^{-1}$ \\
    Boltzmann constant  & $k_B$ & 1.38 $\times 10^{-23}$ J K$^{-1}$ \\
    Stefan-Boltzmann constant & $\sigma$ & 5.7 $\times 10^{-8}$ W m$^{-2}$
K$^{-4}$ \\
    Gravitational constant & $G$ & 6.7 $\times$ 10$^{-11}$ m$^3$ kg$^{-1}$
s$^{-2}$ \\
    \br
  \end{tabular}
  \end{indented}
\end{table}

\begin{table}
  \caption{\label{Table:PM acceleration disturbances}The estimated values
for the proof-mass acceleration disturbances at the frequency of 0.1 mHz.
  PM and rss denote proof mass and root-sum-square, respectively. The values
are in units of 10$^{-15}$ m s$^{-2}$ Hz$^{-1/2}$.
  See sections \ref{Anp} and \ref{back-action}, and tables
  \ref{Table:parameter values} and \ref{Table:Physical constants}
  for notation.}
\begin{indented}
\item[]\begin{tabular}{@{}llll}
\br
  [$10^{-15}$ m s$^{-2}$ Hz$^{-1/2}$] & & Expressions & ASTROD I\\
    \mr
  \multicolumn{4}{l}{(a) PM environmental acceleration disturbances}  \\
  $f_{m1}$ & Magnetic & $\frac{2 \chi_m}{\rho \mu_0
\xi_m}|\vec{\nabla}B_{sc}|\delta B_{sc}$ & 1.2  \\
  $f_{m2}$ & Magnetic & $\frac{\sqrt{2} \chi_m}{\rho \mu_0
\xi_m}|\vec{\nabla}B_{sc}|\delta B_{ip}$ & 3.4  \\
  $f_{m3}$ & Magnetic & $\frac{1}{\sqrt{2} m_p}|\vec{M_r}||\vec{\nabla}
(\delta B)|$ & 1.6 \\
  $f_{L1}$ & Magnetic & $\frac{v}{\xi_e m_p}\delta q B_{ip}$ & 0.0017  \\
  $f_{L2}$ & Magnetic & $\frac{v}{\xi_e m_p}q \delta B_{ip}$ & 0.91 \\
  $f_c$ & Cosmic rays & $\sqrt {\frac{4mE_d \lambda}{m_p^2} }$ & 0.0015 \\
  $f_{rg}$ & Residual gas & $2\sqrt {\frac{P A_p}{m_p^2}}(3 k_B T_p
m_N)^{1/4}$ & 0.74 \\
  $f_{re}$ & Radiometric & $\frac{A_p P}{2m_p}\frac{\delta T_d}{T_p}$ & 0.17
  \\
  $f_{og}$ & Outgassing & 10$f_{re}$ & 1.7 \\
  $f_{tr}$ & Thermal radiation & $\frac{8}{3}\frac{\sigma A_p}{m_p c}T_p^3
\delta T_d$ & 0.13 \\
  $f_{gg}$ & Gravity gradients & $\frac{2 G M_{dis}}{x^2}CTE\cdot\delta
T_{sc}$ & 2.7 \\
  \multicolumn{3}{l}{rss($f_{m1}-f_{gg}) \equiv f_{nep} $} & 5.2 \\
  \mr
  \multicolumn{4}{l}{(b)PM sensor back-action acceleration disturbances}\\
  $f_{b1}$ & $\delta V_d \times V_{0g}$ & $\frac{C_x}{m_p
d}\frac{C_g}{C}V_{0g}\delta V_d$ & 1.4 \\
  $f_{b2}$ & $\delta V_d \times q$ & $\frac{1}{m_p d}\frac{C_x}{C} q \delta
V_d$ & 4.8 \\
  $f_{b3}$ & $\delta q \times V_{d}$ & $\frac{1}{m_p d}\frac{C_x}{C} V_d
\delta q$ & 2.9 \\
  $f_{b4}$ & $\delta q \times q$ & $\frac{1}{m_p
d}\frac{C_x}{C^2}\frac{\Delta d}{d} q \delta q$ & 0.040 \\
  $f_{ba}$ & Readout electronics & (see table 1 of \cite{Cavalleri2001}) &
1.8 \\
  $f_{pe}$ & Patch fields & $\frac{1}{m_p d} \frac{C_x}{C} V_{pe} \delta q$
& 29 \\
  $f_{dl}$ & Dielectric losses & $\sqrt{2} \frac{C_x}{m_p d} V_{0} \delta
  v_{diel}$& 1.6 \\
  \multicolumn{3}{l}{rss($f_{b1}-f_{dl}) \equiv f_{nbp} $} & 30 \\
  \multicolumn{3}{l}{rss($f_{m1}-f_{dl}) \equiv f_{np} $} & 30  \\
  \br
\end{tabular}
\end{indented}
\end{table}

\subsection{\textit{Magnetostatic interaction}}

The lowest order force on a proof mass with a magnetic moment
$\vec{M_p}$ in an external magnetic field $\vec{B}$ is given by
$\vec{F_m}$ = $\vec{\nabla}(\vec{M_p}\cdot\vec{B})$
\cite{Jackson1998}. The magnetic moment of the proof mass is a
vector sum of the remanent moment $\vec{M_r}$ and the induced moment
\cite{Hanson2003}: $\vec{M_p} = \vec{M_r}+(\chi_m
V_p/\mu_0)\vec{B}$, where $\chi_m$ and $V_p$ are magnetic
susceptibility and the volume of the proof mass, respectively;
$\mu_0 = 1.26 \times 10^{-6}$\,N\,A$^{-2}$ is the permeability of
vacuum. The external magnetic field would be given by the
superposition of the interplanetary magnetic field $\vec{B_{ip}}$
and a local magnetic field $\vec{B_{sc}}$. Dominant terms in
acceleration of the proof mass due to the induced magnetic moment
produce acceleration disturbances noted $f_{m1}$ and $f_{m2}$ in
table \ref{Table:PM acceleration disturbances}, where $\xi_m$ is a
scaling factor for possible suppression by magnetic shielding.

The average interplanetary magnetic field at 1\,AU from the Sun
varies from $10^{-9}$ to 3.7 $\times$ 10$^{-8}$\,T \cite{Cambridge}.
Ulysses data obtained near 1\,AU from the Sun (figure 9 of
\cite{Balogh1992}) showed a $\nu^{-2/3}$ dependence of the variation
in the interplanetary magnetic field and $\delta B_{ip}$ can be
inferred to be about $10^{-7}$\,T\,Hz$^{-1/2}$ at 0.1\,mHz. As the
behavior of $\delta B_{ip}$ at 0.5\,AU from the Sun is uncertain, we
use a somewhat higher value of $\delta B_{ip} = 4 \times
10^{-7}$\,T\,Hz$^{-1/2}$ (see footnote in section \ref{st:Lorentz}).
According to the formulation studies for LISA and the implementation
work of the LISA Pathfinder, batteries and micro-thrusters are the
principal suspects of the origins of the local magnetic field in
LISA\footnote{This is a comment from one of the referees.}. We need
elaborate modeling works to estimate the magnitude of local magnetic
field. Here, we use the same values used in analyses for LISA
\cite{Schumaker2003,Shaul2005}: $B_{sc} \approx 8 \times
10^{-7}$\,T, $\delta B_{sc}$ = 10$^{-7}$\,T\,Hz$^{-1/2}$ and
$|\vec{\nabla} B_{sc}|$ $\approx$ 3 $B_{sc}r_m^{-1}$ $\approx$ 3
$\times$ 10$^{-6}$\,T\,m$^{-1}$ (where $r_m$ = 0.75\,m). Silvestri
et al.\ have reported that the magnetic susceptibility of five
samples ranged from $-$2.8 $\times$ 10$^{-5}$ to $-$2.1 $\times$
10$^{-5}$ for two of them without traceable iron contamination and
from +1.1 $\times$ 10$^{-5}$ to +8.8 $\times$ 10$^{-5}$ for the rest
samples with a trace of iron \cite{Silvestri2003}. Lower
susceptibility may be achievable by controlling the manufacturing
process of the alloy. The requirement for LISA is $\chi_m$ = 3
$\times$ 10$^{-6}$ \cite{Stebbins2004}. We use a somewhat moderate
value of $\chi_m$ = 5 $\times$ 10$^{-5}$. Assuming $\xi_m$ = 1 as
per \cite{Schumaker2002,Schumaker2003}, we obtain $f_{m1}$ = 1.2
$\times$ 10$^{-15}$\,m\,s$^{-2}$\,Hz$^{-1/2}$ and $f_{m2}$ = 3.4
$\times$ 10$^{-15}$\,m\,s$^{-2}$\,Hz$^{-1/2}$ at 0.1\,mHz.

Acceleration disturbance due to the magnetic remanent moment is
given by $f_{m3}$ listed in table \ref{Table:PM acceleration
disturbances} (equation 4(c) of \cite{Hanson2003}), where
$\Delta(\delta B)$ is time-varying magnetic field gradients.
According to measurements by Gill et al., magnetic remanent moment
of three 4.2-cm 70/30 Au-Pt alloy cubes (one of them was 73/27 Au-Pt
alloy) ranged from 3.5 $\times$ 10$^{-9}$\,A\,m$^2$\,kg$^{-1}$ to
3.1 $\times$ 10$^{-8}$\,A\,m$^2$\,kg$^{-1}$ \cite{Gill2002}. By
scaling the largest measured remanent moment by weight, it is $5.4
\times 10^{-8}$\,A\,m$^2$ for a 1.75-kg proof mass. By using a
somewhat relaxed value of 1 $\times$ 10$^{-7}$\,A\,m$^2$ and
$|\nabla (\delta B)| = 4 \times 10^{-8}$\,T\,m$^{-1}$\,Hz$^{-1/2}$
\cite{Hanson2003}, which is a factor of four higher than the assumed
requirement for LISA in \cite{Stebbins2004}, we obtain $f_{m3} = 1.6
\times 10^{-15}$\,m\,s$^{-2}$\,Hz$^{-1/2}$ at 0.1\,mHz. In the error
estimates for LISA \cite{Stebbins2004}, magnetic remanent moment of
2 $\times$ 10$^{-8}$\,A\,m$^2$ is used.

\subsection{\textit{Lorentz force}}\label{st:Lorentz}

The proof mass in orbit would be charged up by cosmic-ray impacts
\cite{Jafry1997,LISA}. As the charged proof mass moves through the
interplanetary magnetic field ($\vec{B_{ip}}$) with a velocity $v$
of about 4 $\times$ 10$^4$\,m\,s$^{-1}$, it experiences the Lorentz
force: $\vec{F_{L}}$ = $q \vec{v} \times \vec{B_{ip}}$, where $q$ is
the built-up charge. Acceleration disturbances due to the
fluctuation of the charge build-up ($\delta q$) in the proof mass
and of the average interplanetary magnetic field are given by
$f_{L1}$ and $f_{L2}$ (see table \ref{Table:PM acceleration
disturbances}), respectively, where $\xi_e$ is an electrostatic
shielding factor; in the rest frame of the proof mass, the motion of
the proof mass through the interplanetary magnetic field generates
an electrostatic field.

By assuming the Poisson distribution of cosmic-ray impacts, the
average charge fluctuation spectral density can be defined as
$\delta q(\omega) \equiv \sqrt{2 e \dot{q}}/\omega$. For the
frequency ($\nu = (2 \pi)^{-1} \omega$) of 0.1\,mHz and the
effective charging rate ($\dot{q}$) of 288 $\mathrm{+e}$\,s$^{-1}$,
which was estimated for a LISA proof-mass (46-mm cube) by a
simulation using GEANT 4 toolkit \cite{Araujo2004}, we obtain
$\delta q(\omega)$ = 6.1 $\times$ 10$^{-15}$\,C\,Hz$^{-1/2}$. Using
this value, $\xi_e$ = 10 and $B_{ip}$ = 1.2 $\times$ 10$^{-7}$\,T,
we obtain $f_{L1}$ = 1.7 $\times$
10$^{-18}$\,m\,s$^{-2}$\,Hz$^{-1/2}$ at 0.1\,mHz. The maximum value
of $B_{ip}$ at 1\,AU is about 3 $\times$ 10$^{-8}$\,T. We
tentatively use four times the maximum value as $B_{ip}$ at 0.5\,AU,
by assuming 1/(distance)$^2$ dependence of $B_{ip}$\footnote{The
solar dipole field at 0.5\,AU is less than $\sim$ 10$^{-10}$\,T. The
main magnetic field at 0.5\,AU to 1\,AU is due to the influence of
solar winds which attenuate with (distance)$^2$.}. Because the
volume of the ASTROD~I proof-mass is about 10\,\% smaller than the
LISA proof mass, the charging rate for ASTROD~I might be smaller
than the value for LISA. Bao et al.\ are working on simulations to
estimate the charging rates for ASTROD~I \cite{Bao2004}. Stebbins et
al.\ use $\xi_e$ = 100, which is one order of magnitude larger than
the value we used here, in the current error estimates for LISA
\cite{Stebbins2004}. Taking a nominal maximum build-up charge $q$ =
10$^{-12}$\,C, which is one order of magnitude larger than the value
used in the error estimates for LISA \cite{Schumaker2002,
Schumaker2003,Stebbins2004}, we obtain $f_{L2}$ = 9.1 $\times$
10$^{-16}$\,m\,s$^{-2}$\,Hz$^{-1/2}$.

\subsection{\textit{Cosmic-ray impacts}}

Some cosmic rays get stopped in the proof mass and deposit momentum
\cite{Jafry1997,LISA}. Assuming the Poisson distribution of cosmic
ray impact, spectral density of momentum transfer ($p$) is $2p^2
\lambda$, where $\lambda$ is the fluctuation in the impact rate
\cite{Jafry1997}. The acceleration disturbance due to the
fluctuation in the impact rate is given by $f_c$, listed in table
\ref{Table:PM acceleration disturbances}. The impact rate was
inferred from simulations done for LISA; by adding the effects of
all stopped particles (protons and helium) and taking into account
their directions, the acceleration disturbance by momentum transfer
was estimated to be $\approx$ 2 $\times$
10$^{-18}$\,m\,s$^{-2}$\,Hz$^{-1/2}$ for a LISA proof mass
\cite{Jafry1997}. This corresponds to a disturbance due to momentum
transfer by protons (mass $m = $1.7 $\times$ 10$^{-27}$\,kg), with
an impact rate of $\sim$ 30\,s$^{-1}$, at incident energy $E_d$ =
200\,MeV (= 3.2 $\times$ 10$^{-11}$\,J). Using these values, the
acceleration disturbance becomes $f_c$ = 1.5 $\times$
10$^{-18}$\,m\,s$^{-2}$\,Hz$^{-1/2}$.

\subsection{\textit{Residual-gas impacts}}

From the kinetic theory, the number of residual-gas molecules
(assumed as ideal gas) that pass an area ($A_p$) of the proof mass
per second is given by $\varpi$ = $n A_P \overline{v}/6$, where $n$
= $P (k_B T_P)^{-1}$ is the number density of the molecules and
$\overline{v}$ = $\sqrt{3 k_B T_P m_N^{-1}}$ is the average thermal
velocity ($P$ is the pressure of residual gas, $k_B$ = 1.38 $\times$
10$^{-23}$\,J\,K$^{-1}$ is the Boltzmann constant, $T_P$ is the
temperature of the proof-mass housing and $m_N$ = 4.65 $\times$
10$^{-26}$\,kg is the mass of nitrogen molecules). Assuming the
Poisson distribution of the impact rate, we define the spectral
density of fluctuations in $\varpi$ as $\delta \varpi(\omega)$
$\equiv$ $\sqrt{2\varpi}$.

Acceleration due to the residual gas impacts is given by $2 m_N
\varpi \overline{v} m_p^{-1}$ and acceleration disturbance due to
fluctuations in the impact rate of residual gas is given by $f_{rg}$
listed in table \ref{Table:PM acceleration disturbances}. For $P$ =
10$^{-5}$\,Pa, $A_P$ = 1.75 $\times$ 10$^{-3}$\,m$^2$ and $T_P$ =
293\,K, we obtain $f_{rg}$ = 7.4 $\times$
10$^{-16}$\,m\,s$^{-2}$\,Hz$^{-1/2}$. Stebbins et al.\ use
3$\times$10$^{-6}$\,Pa as the residual gas pressure around the proof
mass in the error estimates for LISA \cite{Stebbins2004}.

\subsection{\textit{Radiometric effect}}

Acceleration disturbance due to the radiometric effect (e.g.
\cite{Blaser1996,Touboul1996,Nobili2001,Worden2001,Rodrigues2003})
is given by $f_{re}$ listed in table \ref{Table:PM acceleration
disturbances}, where $\delta T_d$ is fluctuation in temperature
difference across the proof mass housing. This value has to be
estimated by carrying out thermal modeling. According to thermal
analysis for LISA, temperature fluctuation on the optical bench, to
which the proof mass housing is mounted, due to power dissipation of
amplifiers is about 3.0 $\times$ 10$^{-5}$\,K\,Hz$^{-1/2}$ at 1\,mHz
(table 6.2-28 of \cite{FTR}). By assuming that the fluctuation rises
as $1/\nu$, the temperature fluctuation of the optical bench is
$\delta T_{ob} = 3.0 \times 10^{-4}$\,K\,Hz$^{-1/2}$ at 0.1\,mHz. At
the frequency of 0.1\,mHz and higher frequencies, the temperature
fluctuation on the optical bench would be dominated by the
fluctuation in the power dissipation \cite{Bender2003}; fluctuation
due to solar irradiance at 0.1\,mHz is $1.1 \times
10^{-6}$\,K\,Hz$^{-1/2}$ (table 6.2-16 of \cite{FTR}). The ratio
$\delta T_{ob}$/$\delta T_{d}$ would range from 30 to 100
\cite{Bender2003}. By using a value of 30 for the ratio, we obtain
$\delta T_d = 1.0 \times 10^{-5}$\,K\,Hz$^{-1/2}$ at 0.1\,mHz. By
using this value, we obtain $f_{re}$ = 1.7 $\times$
10$^{-16}$\,m\,s$^{-2}$\,Hz$^{-1/2}$ at 0.1\,mHz.

\subsection{\textit{Temperature dependent outgassing effect}}

Outgassing from walls of the sensor cage is thought to produce
greater acceleration noise than the radiometric effect
\cite{Ruediger2002, Dolesi2003}. An analysis done for the LISA
Technology Package (LTP), assuming a simple model of flow circuit
with a linear approximation, shows that the outgassing effect is
nearly 10 times the radiometric effect \cite{Dolesi2003}. By using
this estimate and the estimate we made for the radiometric effect in
the previous section, we obtain $f_{og}$ = 1.7 $\times$
10$^{-15}$\,m\,s$^{-2}$\,Hz$^{-1/2}$ at 0.1\,mHz for ASTROD~I.

\subsection{\textit{Thermal radiation pressure}}

By assuming a perfectly reflecting surface of the proof mass,
thermal radiation pressure produces acceleration disturbance
($f_{tr}$ in table \ref{Table:PM acceleration disturbances}) due to
fluctuations in the temperature difference across the proof-mass
housing. In the expression of $f_{tr}$ listed in table \ref{Table:PM
acceleration disturbances}, $\sigma$ = 5.7 $\times$
10$^{-8}$\,W\,m$^{-2}$\,K$^{-4}$ is the Stefan-Boltzmann constant,
$A_p$ is the area of the proof mass and $T_p$ is the temperature of
the proof-mass housing. A factor of one-third is multiplied, as done
in the estimation for LISA by Schumaker
\cite{Schumaker2002,Schumaker2003}, as a margin for the fact that
not all of the radiation momentum is normally incident on the proof
mass. For the housing temperature of 293\,K and the temperature
fluctuation of 1.0 $\times$ 10$^{-5}$\,K\,Hz$^{-1/2}$, we obtain
$f_{tr}$ = 1.3 $\times$ 10$^{-16}$\,m\,s$^{-2}$\,Hz$^{-1/2}$ at
0.1\,mHz. The same value was used for the temperature fluctuation in
the error estimates for LISA \cite {Stebbins2004}.

\subsection{\textit{Gravity gradients due to thermal distortion of
the spacecraft}}

The temperature fluctuation in solar irradiance would cause
fluctuation in distortions of the spacecraft: $\delta x x^{-1} = CTE
\cdot |\delta T_{sc}|$, where $CTE$ and $\delta T_{sc}$ are the
coefficient of thermal expansion and temperature fluctuation of the
spacecraft, respectively. The inherent fluctuation in solar
radiation is $\delta W_0/W_0 = 4 \delta T/T_p \approx 2.8 \times
10^{-3}$\,Hz$^{-1/2}$ at 0.1\,mHz \cite{Pap1999}. Therefore, the
temperature fluctuation is 0.2\,K\,Hz$^{-1/2}$ for $T_p$ = 293\,K.
The gravitational disturbance by a 1-kg mass ($M_{dis}$) separated
from the proof mass in the sensitive axis by $x$ = 0.5\,m of
aluminium structure would be $f_{gg} = 2.7 \times
10^{-15}$\,m\,s$^{-2}$\,Hz$^{-1/2}$ for $\delta T_{sc}$ =
0.2\,K\,Hz$^{-1/2}$ and $CTE$ (of Aluminium) = 2.5 $\times$
10$^{-5}$\,K$^{-1}$. The same disturbing mass ($M_{dis}$ = 1\,kg and
$x$ = 0.5\,m) is assumed in
\cite{Schumaker2002,Schumaker2003,Stebbins2004}. In reality, the
mass to be involved in thermal distortion would be much larger, but
the influence would be largely canceled because of the axial
symmetry in the original spacecraft geometry. The inherent
fluctuation in solar radiation could be reduced largely by thermal
shielding. In the estimation for LISA, $\delta T_{sc}$ =
0.004\,K\,Hz$^{-1/2}$ is used \cite{Schumaker2003,Stebbins2004}. For
a more accurate estimate, gravity effects by thermal and non-thermal
distortions of the spacecraft and the payload have to be studied by
appropriate modeling.

\section{Proof-mass sensor back-action acceleration
disturbances}\label{back-action}

The total mechanical energy of the capacitive sensing system can
be expressed as (e.g. equation (A.3) of \cite{Schumaker2002}):
\begin{equation}
    W = - \frac{1}{2}\sum_i C_i (V_i - V_s)^2 + \frac{1}{2}
    \frac{q^2}{C} + q V_s\label{eq:total mechanical energy}
\end{equation}
where $q$ is a net charge of the proof mass; $C$ is the sum of the
capacitances due to the applied voltages on the surrounding
electrodes $i$ and the potential to ground $g$ : $C = \sum_i C_i
$, where $ i = x_1, x_2, y_1, y_2, z_1, z_2, g$; $V_s$ is the
voltage induced on the proof mass by the applied voltages on the
electrodes: $V_s \equiv C^{-1}\sum_i V_i C_i $. The first term of
$W$ is the total energy done on the proof mass by the applied
voltages of the surrounding electrodes. The second term is the
energy acquired on the proof mass by the image charge on the
surrounding electrodes. The third term is the energy stored on the
proof mass by the deposit of the free charge on the proof mass.

The $x$-component of the force on the proof mass is given by
differentiation of equation (\ref{eq:total mechanical energy}) with
respect to $x$. For a simplicity, we assume that neither the free
charge $q$ nor the potentials $V_i$ have appreciable gradients along
the $x$-axis, as per \cite{Schumaker2002}. Acceleration disturbances
due to fluctuations in the applied voltages and charge can be given
by $f_{b1}$ - $f_{b4}$ as listed in table \ref{Table:PM acceleration
disturbances}. A detailed description on the deviations of $f_{b1}$
- $f_{b4}$ is given in Appendix A of \cite{Schumaker2002}. The
parameter values used in this section and the estimated values are
listed in tables \ref{Table:parameter values} and \ref{Table:PM
acceleration disturbances}, respectively.

To simplify the analysis, several assumptions were made in the
process of deriving the expressions for the four classes of
acceleration disturbances ($f_{b1}$ - $f_{b4}$)
\cite{Schumaker2002}: (1) $C_i$ are all comparable with each other
in magnitude and on the order of $C_x$ $\approx$ 6\,pF (=
$\epsilon_0 A d^{-1}$, where $\varepsilon_0$ = 8.9 $\times$
10$^{-12}$\,F\,m$^{-1}$ is the permittivity of vacuum, $A$ is the
area of each electrode and $d$ = 2\,mm is the gap); (2) only the
capacitances $C_{x1}$ and $C_{x2}$ have nonzero gradients along the
$x$-axis: $C'_{x1}$ = $-C'_{x2}$, and the gradient of the total
capacitance $C$ $\approx$ 6 $C_x$ is: $C'$ $\approx$ $(C_x\Delta d)
d^{-2}$, where $\Delta d$ is the gap asymmetry in the $x$-direction;
(3) the average of the potentials on opposing faces is same for all
three axes and expressed as $V_{x_0}$ ($\equiv (V_{x1} + V_{x2})/2 =
$ 0.1\,V); (4) the magnitude of the fluctuation in potential $\delta
V_i$ is all identical and take the value of the average fluctuation
of all the potentials for the three axes and the voltage to ground.
We express the fluctuation as $\delta V_{x_0}$. It should be noted
that we do not consider any cross-talk effects that arise in the
sensitive axis due to forces applied to the other degrees of
freedom.

\subsection{Fluctuations in voltage imbalance and charge}

$f_{b1}$ is associated only with sensing voltages but not the free
charge, and its value is $f_{b1}$ $\approx$  1.4 $\times$
10$^{-15}$\,m\,s$^{-2}$\,Hz$^{-1/2}$, by assuming $V_{0g}$ $\equiv$
$V_{x_0}-V_g$ = 0.05\,V. This value is larger than the value used
for LISA \cite{Stebbins2004} by a factor of five. Also, we use
$\delta V_d$ = 1.0 $\times$ 10$^{-4}$\,V\,Hz$^{-1/2}$ as fluctuation
in voltage imbalance $V_d$ ($\equiv$ $V_{x1} - V_{x2}$ = 0.01\,V).
This value ($\delta V_d$) is one order of magnitude larger than the
value used for LISA\cite{Schumaker2002,Schumaker2003,Stebbins2004}.
Further, we assume that $C_g/C$ $\approx$ 1/6, as per
\cite{Schumaker2002,Schumaker2003}.

$f_{b2}$ and $f_{b3}$ arise from fluctuations in the force due to
the interaction between the net free charge and applied sensing
voltages. $f_{b2}$ is due to fluctuation in voltage imbalance and
its value is $f_{b2}$ $\approx$ $-$4.8 $\times$
10$^{-15}$\,m\,s$^{-2}$\,Hz$^{-1/2}$, where the net free charge on
the proof mass is set to the nominal maximum built-up charge,
described earlier. $f_{b3}$ is due to fluctuation in charge and its
value is $f_{b3}$ $\approx$ $-$2.9 $\times$
10$^{-15}$\,m\,s$^{-2}$\,Hz$^{-1/2}$. The value of $V_d$ used here
is larger than the one used in the error estimates for LISA
\cite{Stebbins2004} by a factor of two. $f_{b4}$ is associated only
with the free charge and its value is $f_{b4}$ $\approx$ 4.0
$\times$ 10$^{-17}$\,m\,s$^{-2}$\,Hz$^{-1/2}$. In the estimation, we
assume that the gap asymmetry in the $x$-direction is $\Delta d$ =
10\,$\mu$m, which is larger than \cite{Schumaker2002,Schumaker2003}
by one order of magnitude.

\subsection{Readout electronics}

Readout electronics for the capacitive sensing will be similar to
ones studied for LISA (e.g.
\cite{Speake1997,Cavalleri2001,Weber2003}). For the standard
resonant inductive-bridge scheme discussed in \cite{Cavalleri2001},
the
sources of disturbances due to readout electronics can be classified into
two categories: imperfections in the
capacitance bridge and the electric noise in the detecting circuit.
The former includes fluctuations of inductance imbalance ($\Delta
L/L$), fluctuations of mutual inductance imbalance ($\Delta M/M$),
bias oscillator relative amplitude noise ($\Delta V/V_{M0}$, where
$V_{M0}$ is the 100\,kHz bias voltage capacitively applied to the
proof mass by injection electrodes on insensitive faces (the $y$
and/or $z$ faces) of the proof mass; the sensing electrodes are
grounded) and bias oscillator phase noise ($\Delta \phi$). The
latter includes current noise and thermal noise.

Using the expressions for these sources of the disturbances given in
table 1 of \cite{Cavalleri2001}, the main contributions to the
acceleration disturbance at 0.1\,mHz for ASTROD~I are from $\Delta
V/V_{M0}$ ($\approx$ 1.5 $\times$
10$^{-15}$\,m\,s$^{-2}$\,Hz$^{-1/2}$) and $\Delta \phi$ ($\approx$
9.1 $\times$ 10$^{-16}$\,m\,s$^{-2}$\,Hz$^{-1/2}$), and the thermal
noise ($\approx$ 3.8 $\times$ 10$^{-17}$\,m\,s$^{-2}$\,Hz$^{-1/2}$)
and the current noise ($\approx$ 1.3 $\times$
10$^{-17}$\,m\,s$^{-2}$\,Hz$^{-1/2}$). Contributions from
fluctuations of $\Delta L/L$ and $\Delta M/M$ are insignificant. In
the estimation, we use the following parameters based on
experimental results reported in \cite{Cavalleri2001}: $\Delta L/L
\approx 10^{-4}$, $\Delta M/M \approx 6 \times 10^{-8}$, $\Delta
V/V_{M0} \approx 10^{-3}$\,Hz$^{-1/2}$, $\Delta \phi \approx 5.7
\times 10^{-4}$\,Hz$^{-1/2}$ and the turn ratio of the transformer
$n$ = 1 (40 turns, inductance of 5\,mH and a quality factor of 165),
and a somewhat relaxed residual imbalance of the bridge $\rho_{dc}
\approx \Delta d/d = 5 \times 10^{-3}$. This relaxation has resulted
in the dominant contributions from the bias oscillator relative
amplitude noise and the phase noise. In \cite{Cavalleri2001}, the
thermal noise is dominant as they use $\rho_{dc} \approx 10^{-4}$ in
their estimation. The rss of these disturbances is $f_{ba}$
$\approx$ 1.8 $\times$ 10$^{-15}$\,m\,s$^{-2}$\,Hz$^{-1/2}$ at
0.1\,mHz and listed in table \ref{Table:PM acceleration
disturbances}.

\subsection{Patch field voltage}
Even when the sensing voltages are not applied on the electrodes,
differences in local surface properties of the electrodes and the
proof mass could lead a potential difference, patch-field voltage,
between them \cite{Speake1996}. The charge fluctuations $\delta q$
result in acceleration disturbance $f_{pe}$ (table \ref{Table:PM
acceleration disturbances}) through the patch field
\cite{Weber2003}. By taking the average patch-filed voltage
difference between opposing electrodes as $V_{pe}$=0.1\,V
\cite{Schumaker2002,Schumaker2003,Weber2003}, we obtain $f_{pe}$ =
2.9 $\times$ 10$^{-14}$\,m\,s$^{-2}$\,Hz$^{-1/2}$ at 0.1\,mHz. This
is the dominant contribution to the total acceleration disturbance
of the ASTROD~I proof mass. The LISA study team is investigating the
possibility of measuring and compensating voltage differences across
capacitors, during the mission commission process, to considerably
better than 0.01\,V \cite{Stebbins2004}.

\subsection{Dielectric losses}

Dielectric losses are thought to stem from surface contamination
of electrodes and produces thermal voltage noise \cite{Vitale1998,
Weber2003}:
\begin{equation}
\delta v_{diel} = \sqrt{4 k_B T_p \frac{\delta}{\omega C_x}}
\end{equation}
where $\delta$ is loss angle. The upper limit of $\delta$ is
reported to be $10^{-5}$ for Al electrodes \cite{Speake1999}. For
$\delta = 10^{-5}$, this voltage noise is about 6.6 $\mu$V
Hz$^{-1/2}$ at 0.1\,mHz and produces acceleration disturbance
$f_{dl}$ (see table \ref{Table:PM acceleration disturbances} for the
expression), through residual dc bias voltage on electrodes, in the
sensitive axis \cite{Vitale1998, Weber2003}. By making the same
assumption of the average potential difference $V_0 = 0.1$\,V
between a given electrode and the proof mass as \cite{Weber2003}, we
obtain $f_{dl}$ $\approx$ 1.6 $\times$
10$^{-15}$\,m\,s$^{-2}$\,Hz$^{-1/2}$. In the error estimates for
LISA \cite{Stebbins2004}, $V_0 = 0.01$\,V is used.

\subsection{Summary of the direct acceleration disturbances of the proof
mass}

By adding in quadrature, the rss of the proof-mass environmental
acceleration disturbances ($f_{nep}$) and the sensor back-action
acceleration disturbances ($f_{nbp}$) are 5.2 $\times$
10$^{-15}$\,m\,s$^{-2}$\,Hz$^{-1/2}$ and 3.0 $\times$
10$^{-14}$\,m\,s$^{-2}$\,Hz$^{-1/2}$, respectively. Therefore, the
total direct proof-mass acceleration disturbance ($f_{np}$) is 3.0
$\times$ 10$^{-14}$\,m\,s$^{-2}$\,Hz$^{-1/2}$ at 0.1\,mHz, dominated
by the sensor back-action acceleration disturbances.

\section{Proof mass-spacecraft coupling}\label{stiff}

The stiffness $K$ is considered to stem from gravity gradients
($K_{gg}$), fluctuations in sensing capacitance and capacitance
gradients ($K_{s1}$, $K_{s2}$ and $K_{s3}$), bias voltage
($K_{s4}$), patch field voltage ($K_{s5}$) and magnetic field
gradients ($K_{m1}$ and $K_{m2}$). Table \ref{Table:Stiffness}
gives a summary of estimated values for these sources of
stiffness. The expressions used in the estimations are briefly
described below. A detailed description on deviations of the
expressions for $K_{gg}$, $K_{s1}$, $K_{s2}$, $K_{s3}$ and $Ks_5$
is given in Appendix A of \cite{Schumaker2002}\footnote{$K_{s5}$
in text is noted as $K_{s4}$ in \cite{Schumaker2002}.}.

\subsection{Gravity gradients}
For a given disturbing point mass ($M_{dis}$) at a distance $x$ from
the center of mass of the proof-mass on the sensitive axis, the
amplitude of the acceleration disturbance of the proof-mass caused
by a positional fluctuation $X_{ps}$ is $ a_{gg} \equiv K_{gg}
X_{ps}$, where $K_{gg}$ is given in table \ref{Table:Stiffness}.

Making the same assumption of $M_{dis}$ = 0.03\,kg and $x$ = 0.05\,m
as \cite{Stebbins2004}, we obtain $K_{gg}$ $\approx$ 3.2 $\times$
10$^{-8}$\,s$^{-2}$. For a more detailed analysis, the
identification of the disturbing mass is necessary. Gravitational
modeling for ASTROD~I is in progress \cite{Shiomi2005}.

This disturbance arises from any positional fluctuation
and is different from the gravity gradient caused by thermal
distortion or motion ($f_{gg}$), which was discussed earlier.

\subsection{Fluctuations in capacitive sensing}
The expressions for $K_{s1}$, $K_{s2}$ and $K_{s3}$ can be obtained
in the similar way as $f_{b1}$, $f_{b2}$, $f_{b3}$ and $f_{b4}$ in
the previous section, under the following assumptions
\cite{Schumaker2002}: fluctuations in the capacitances $C_{x1}$ and
$C_{x2}$ and their derivatives produce disturbing forces in the
$x$-direction, but not the fluctuations in other capacitances or
their derivatives; we ignore the cross-coupling effects. Also, we
assume $\delta C_{x1}$ = $- \delta C_{x2} \approx (C_x/d) \delta x$
and $\delta C'_{x1}$ = $\delta C'_{x2} \approx (C_x/d^2) \delta x$.

$K_{s1}$ is due to the fluctuations in the Coulomb interaction
between the charged proof mass and the image charges on the
surrounding electrodes. $K_{s2}$ arises from interaction between
the net free charge $q$ on the proof-mass and the average
electrode voltages. $K_{s3}$ is due to the applied voltages across
electrodes and the voltage difference across opposite electrodes.
The estimated values for $K_{s1}$, $K_{s2}$ and $K_{s3}$ are given
in table \ref{Table:Stiffness}.

\subsection{Bias voltage}
By assuming that the sensing electrodes are grounded, the dominant
term of the readout stiffness along the sensitive axis is given by
$K_{s4}$ \cite{Weber2003}, listed in table \ref{Table:Stiffness}.
Assuming, as per \cite{Weber2003}, that the proof mass is biased to
$V_{M0} = 0.6$\,V, $K_{s4}$ $\approx$ 3.1 $\times$
10$^{-7}$\,s$^{-2}$.

\subsection{Patch field voltage}
By assuming a nominal patch-field voltage of $V_{pe}$ = 0.1\,V
\cite{Schumaker2002,Schumaker2003,Weber2003} and an overall
multiplicative factor of $\gamma$ = 5 as per
\cite{Schumaker2002,Schumaker2003}, the contribution due to the
patch field to the proof mass-spacecraft coupling is $K_{s5}$
$\approx$ 1.2 $\times$ 10$^{-9}$\,s$^{-2}$ (table
\ref{Table:Stiffness}).

\subsection{Magnetic field gradients}

The magnetic stiffness is given by $K_{m}$ = $\frac{1}{m_p}
\vec{\nabla} [\vec{\nabla} (\vec{M_p} \cdot \vec{B})]$
\cite{Hanson2003}. The expressions for dominant terms of the
stiffness due to the induced magnetic moment and the remanent moment
are given by $K_{m1}$ and $K_{m2}$ (as listed in table
\ref{Table:Stiffness}), respectively, where $|\vec{\nabla}^2 B_{sc}|
\approx 12 B_{sc}r_m^{-2}$ = 1.7 $\times$ 10$^{-5}$\,T\,m$^{-2}$ for
$r_m$ = 0.75\,m. Their contributions to the total stiffness are
insignificant.

\subsection{Summary of the estimated values of stiffness}

These contributions to the coupling constant $K$ are summarized in
table \ref{Table:Stiffness}. The rss of the coupling constant is 3.1
$\times$ 10$^{-7}$\,s$^{-2}$. This is slightly below the requirement
for the total stiffness in LISA (4 $\times$ 10$^{-7}$\,s$^{-2}$
\cite{Weber2003}).

\begin{table}
\caption{\label{Table:Stiffness}The proof-mass stiffness in units of
10$^{-9}$\,s$^{-2}$ (see sections \ref{back-action} and \ref{stiff},
and tables \ref{Table:parameter values} and \ref{Table:Physical
constants} for notation.)}
  \begin{indented}
  \item[]\begin{tabular}{@{}llll}
    \br
     [$10^{-9}$ $\mathrm{s}^{-2}$] & & \hspace{-3mm} Expressions &
\hspace{-3mm} ASTROD I \\
     \mr
    $K_{gg}$ &\hspace{-3mm} Gravity gradients & $\hspace{-2mm} 2\frac{G
M_{dis}}{x^3}$ & \hspace{-3mm} 32 \\
    $K_{s1}$ & \hspace{-3mm} Image charges & $\hspace{-2mm} \frac{1}{m_p C
d^2}\left(\frac{C_x}{C}\right)q^2$ & \hspace{-3mm} 0.66  \\
    $K_{s2}$ & \hspace{-3mm} $q \times V_{0g}$ & \hspace{-2mm}
$\left(\frac{2}{m_p d^2}\right)\left(\frac{C_x}{C}\right)
    \left(\frac{C_g}{C}\right)q V_{0g}$
     &\hspace{-3mm} 0.40 \\
    $K_{s3}$ & \hspace{-3mm} Applied voltages & \hspace{-2mm}
$\frac{C_x}{m_p d^2}\left
\{\left(\frac{C_x}{C}+\frac{1}{4}\right)V_{d}^2+\left(\frac{C_g}{C}\right)^2
V_{0g}^2 \right \}$
    &\hspace{-3mm} 0.095 \\
    $K_{s4}$ & \hspace{-3mm} Bias voltage & \hspace{-2mm} $\frac{C_x}{m_p
d^2}V_{M0}^2$
    &\hspace{-3mm} 3.1 $\times$ 10$^{2}$ \\
    $K_{s5}$ & \hspace{-3mm} Patch fields & \hspace{-2mm} $\gamma
\left(\frac{C_x}{m_p d^2}\right)\left(\frac{C_g}{C}\right)^2 V_{pe}^2$
    &\hspace{-3mm} 1.2 \\
    $K_{m1}$ & \hspace{-3mm} Induced magnetic moments & $\frac{2\chi_m}{\rho
\mu_0} \left\{|\vec{\nabla}
    B_{sc}|^2 + B_{sc}|{\vec{\nabla}}^2 B_{sc}|\right\}$
    &\hspace{-3mm} 9.0 $\times$ 10$^{-5}$ \\
    $K_{m2}$ & \hspace{-3mm} Magnetic remanent moments &
$\frac{1}{\sqrt{2}m_p}|\vec{M_r}||{\vec{\nabla}}^2
    B_{sc}|$ &\hspace{-3mm} 6.9 $\times$ 10$^{-4}$ \\
    \mr
    \multicolumn{3}{l}{rss($K_{gg}$ - $K_{m2}$) $\equiv$ $K$} &
\hspace{-2mm} 3.1 $\times$ 10$^{2}$ \\
    \br
  \end{tabular}
  \end{indented}
\end{table}

\section{Requirements for the readout sensitivity and spacecraft
control-loop
gain}\label{st:requirements}

We have estimated values for the coupling constant $K$, the direct
spacecraft acceleration disturbance $f_{ns}$ and the direct
proof-mass acceleration disturbance $f_{np}$. By using the
expression for the total acceleration disturbance of $f_p$ (equation
(\ref{eq:Control-model})), we infer the requirements for the readout
sensitivity $X_{nr}$ and the spacecraft control-loop gain $u$. In
this process, we allocate an identical magnitude $f_a$ to each term
of the expression; $f_a^2 = f_p^2/3$. For ASTROD~I, the noise goal
is $10^{-13}$\,m\,s$^{-2}$\,Hz$^{-1/2}$ at 0.1\,mHz and, therefore,
$f_a = 5.8 \times 10^{-14}$\,m\,s$^{-2}$\,Hz$^{-1/2}$. The second
term of the expression is $f_{np}$, which was estimated to be 3.0
$\times$ 10$^{-14}$\,m\,s$^{-2}$\,Hz$^{-1/2}$ (table \ref{Table:PM
acceleration disturbances}); it is about a factor of 2 smaller than
the allocated requirement. With the estimated total stiffness $K =
3.1 \times 10^{-7}$\,s$^{-2}$, we obtain $X_{nr} \leq 1.9 \times
10^{-7}$\,m\,Hz$^{-1/2}$ from the first term and $u \geq 3.8 \times
10^5$ from the last term of the expression.

From figure \ref{Fig:Curve}, one can see that the acceleration noise
spectral density requirements for ASTROD I take its lowest value of
about 0.4 $\times$ 10$^{-13}$\,m\,s$^{-2}$\,Hz$^{-1/2}$ at 0.3\,mHz.
At this frequency, $f_a$ becomes 2.3 $\times$
$10^{-14}$\,m\,s$^{-2}$ Hz$^{-1/2}$. Therefore, the requirement for
$X_{nr}$ becomes more stringent at 0.3\,mHz: $X_{nr} \leq 7.4 \times
10^{-8}$\,m\,Hz$^{-1/2}$. As for the second term of $f_{p}$,
$f_{np}$ is smaller at higher frequencies \cite{Schumaker2002,
Schumaker2003}. Our estimate of $f_{np}$ is dominated by the
contribution from $f_{pe}$, which scales as $\nu^{-1}$. Therefore,
at 0.3\,mHz, $f_{np}$ would be $\sim$ 1 $\times$
10$^{-14}$\,m\,s$^{-2}$\,Hz, which is smaller than $f_{a}$ at
0.3\,mHz. The last term scales as $\omega^{-2}$, and $f_{ns}$ is
expected to be smaller at higher frequencies because of the $\nu
^{-1/3}$ dependence of the fractional fluctuation in solar
irradiance (figure 6 of \cite{Pap1999}). Therefore, by assuming the
same level of the thruster noise, the requirement for $u$ at
0.3\,mHz would be less stringent than that at 0.1\,mHz by a factor
of $\sim$ 4.

In summary, the requirements for the readout sensitivity and the
control loop gain for ASTROD I are $X_{nr} \leq 7.4 \times
10^{-8}$\,m\,Hz$^{-1/2}$ and $u \geq 3.8 \times 10^5$, respectively.

\section{Comparison with LISA}\label{comparison with LISA}

Main relaxed parameter-values are listed in table
\ref{table:comparison with LISA} in comparison with LISA. The values
for LISA are quoted from the current error estimates by Stebbins et
al.\ \cite{Stebbins2004}, except the thruster noise quoted from
\cite{LISA}. Recently the LISA requirements for the thruster noise
and the residual gas pressure have been relaxed further from the
values given in table \ref{table:comparison with LISA} (this is a
comment from one of the referees).
\begin{table}[h]
  \caption{\label{table:comparison with LISA}Relaxed parameter values in
comparison with LISA}
    \begin{indented}
    \item[]\begin{tabular}{@{}lll}
  \br
   $\nu$ = 0.1 mHz & ASTROD I & LISA \\
   \mr
   Maximum charge build-up: $q$ [C] & $10^{-12}$ & $10^{-13}$ \\
   Magnetic susceptibility: $\chi_m$ & $5 \times 10^{-5}$ & $3 \times
10^{-6}$ \\
   Magnetic remanent moment: $|\vec{M_r}|$ [A m$^2$] & 1 $\times$ 10$^{-7}$
& 2 $\times$ 10$^{-8}$\\
   Residual gas pressure: $P$ [Pa] & $10^{-5}$ & ${3 \times 10^{-6}}^{\flat}$ \\
   Electrostatic shielding factors: $\xi_e$ & 10 & 100 \\
   Thruster noise [$\mu$N Hz$^{-1/2}$] & 10 & a few \\
   Fluctuation of temperature in SC: $\delta T_{sc}$ [K Hz$^{-1/2}$] & 0.2 &
0.004 \\
   Voltage difference between average voltage across &&\\
   \hspace{1pt} opposite faces and voltage to ground: $V_{0g}$ [V] & 0.05 &
0.01 \\
   Voltage difference between opposing faces: $V_d$ [V] & 0.01 & 0.005 \\
   Fluctuation of voltage difference across opposite faces: & & \\
   \hspace{1pt} $\delta V_d$ [V Hz$^{-1/2}$] & $10^{-4}$ & $10^{-5}$ \\
   Residual dc bias voltage on electrodes: $V_0$ [V] & 0.1 & 0.01 \\
   \br
  \end{tabular}
  \item[] $^{\flat}$ According to one of the referees, this LISA requirement has been relaxed to 10$^{-5}$ Pa.
  \end{indented}
\end{table}

\section{Summary, discussion and conclusions}

We have estimated the spacecraft acceleration disturbance $f_{ns}$
(section \ref{Ans}), the proof-mass acceleration disturbances
$f_{np}$ (sections \ref{Anp} and \ref{back-action}) and the
stiffness $K$ between the spacecraft and the proof mass (section
\ref{stiff}). By using the expression (\ref{eq:Control-model}), we
have inferred the requirements for the readout sensitivity
$X_{nr}$ and the control-loop gain $u$ (section
\ref{st:requirements}).

Table \ref{Table:Disturbances and requirements} provides a summary
of the estimated acceleration disturbances and the stiffness
(section (b)), and the requirements (section (c)). The estimated
total acceleration disturbance $f_p$ at 0.1\,mHz (section (a) of
table \ref{Table:Disturbances and requirements}) is about 13 \% less
than the noise goal of $10^{-13}$\,m\,s$^{-2}$\,Hz$^{-1/2}$. We have
compared the parameter values used in the estimation with LISA in
section \ref{comparison with LISA}.

\begin{table}
\caption{\label{Table:Disturbances and requirements}Acceleration
disturbances and requirements.
PM and SC denote proof-mass and spacecraft, respectively.}
\begin{indented}
\item[]\begin{tabular}{@{}ll}
  \br
                            & ASTROD I \\
  \mr
  \multicolumn{2}{l}{(a)Estimated total acceleration disturbance (at 0.1
mHz):} \\
  $f_p$ [m s$^{-2}$ Hz$^{-1/2}$] & 8.7 $\times$ 10$^{-14}$ \\
  \mr
  \multicolumn{2}{l}{(b)Estimated contributions to $f_p$ (at 0.1 mHz):} \\
   & \hspace{-3 cm} $f_p \approx
X_{nr}(-K)+f_{np}+(f_{ns}+TN_t)K\omega^{-2}u^{-1}$ \\
  Direct PM acceleration disturbances: & \\
  $f_{np}$ [m s$^{-2}$ Hz$^{-1/2}$]& 3.0 $\times$ $10^{-14}$ \\
  Direct SC acceleration disturbances: &  \\
  $f_{ns}+TN_t$ [m s$^{-2}$ Hz$^{-1/2}$]& 2.8 $\times$ $10^{-8}$  \\
  PM-SC spring constant: & \\
   $K$ [s$^{-2}$] & 3.1 $\times$ $10^{-7}$ \\
  \mr
  \multicolumn{2}{l}{(c)Inferred requirements} \\
  Position readout noise: $X_{nr}$ [m Hz$^{-1/2}$]  & 7.4 $\times$ $10^{-8}$
\\
  Control-loop gain: $u$ & 3.8 $\times$ $10^{5}$ \\
    \br
\end{tabular}
\end{indented}
\end{table}

The total direct acceleration disturbance of the proof mass
($f_{np}$) at 0.1\,mHz was estimated to be nearly a factor of two
smaller than the requirement. This $\sim$ 50\,\% margin may be
allocated for unknown disturbances or disturbances that would arise
but have not been studied yet. These unestimated disturbances would
be originated from, for instance, cross-talks in the capacitive
sensing and magnetic damping of the proof mass. An estimate of
acceleration disturbance due to magnetic damping for LISA is about 2
$\times$ 10$^{-16}$\,m\,s$^{-2}$\,Hz$^{-1/2}$ at 0.1\,mHz
\cite{Stebbins2004}. The contribution from the magnetic damping
effect to ASTROD~I would be in the similar order and insignificant.

The sensor back-action acceleration disturbances can be reduced by
increasing the magnitude of the gap $d$. The total stiffness $K$
would be also reduced, for example, by a factor of four by changing
the gap to 4\,mm. The optimum design for the capacitive sensing is
to be discussed based on results from the ongoing laboratory torsion
balance experiment for ASTROD~I \cite{Zhou2005}.

Parameter values we used in this paper are mainly based on the
results of studies done for LISA and LISA Pathfinder. This may be
sufficient for the preliminary estimation. More accurate estimation
would be obtained by carrying out the following works dedicated for
ASTROD~I: (a) modeling local magnetic fields of the spacecraft, (b)
estimating the effective charging rate of the proof-mass, (c)
estimating the cosmic-ray impact rate of the proof-mass, (d)
estimating the micrometeorite impact effects, (e) thermal modeling
of the proof-mass housing and the spacecraft, (f) gravitational
modeling that includes thermal and non-thermal deformation of the
spacecraft and the payload, (g) electrostatic modeling for the
capacitive sensors and (h) estimating environmental factors (such as
the interplanetary magnetic field, solar wind, solar radiation and
cosmic rays) in the varying orbit (0.5\,AU to 1\,AU). Simulations to
estimate charging rates for ASTROD~I are in progress \cite{Bao2004}.

We have tentatively estimated acceleration disturbances for
ASTROD~I. This work has allowed us to set preliminary requirements
for ASTROD~I. To improve the current estimation, the disturbances
that have not been studied yet have to be included and more detailed
modeling works are necessary for ASTROD~I. In comparison with LISA,
requirements for ASTROD~I can be largely relaxed. This will make the
technological developments for ASTROD~I less demanding to meet the
drag-free requirements.

\ack This work was funded by the National Science Council and the
Foundation of Minor Planets of Purple Mountain Observatory. We thank
A. R\"{u}diger, S. Vitale, D. K. Gill, A. Pulido Pat\'{o}n, and,
especially, the referees for useful information on acceleration
disturbances and helpful comments on the manuscript.

\end{document}